\documentclass[aps,prl,twocolumn,superscriptaddress,groupedaddress]{revtex4} 
\usepackage{amsmath}
\usepackage{amssymb}
\usepackage{graphicx}
\usepackage{bm}

\begin{document}

\title{Shaping Electromagnetic Fields}
\author{Bo Liu}
\email{bliu@physics.harvard.edu}
\affiliation{Department of Physics, Harvard University, Cambridge, Massachusetts 02138, USA}
\affiliation{Institute for Applied Computational Science, Harvard University, Cambridge, Massachusetts 02138, USA}
\affiliation{Center for Integrated Quantum Materials, Cambridge, Massachusetts 02138, USA}
\author{Eric J. Heller}
\affiliation{Department of Physics, Harvard University, Cambridge, Massachusetts 02138, USA}
 \affiliation{Center for Integrated Quantum Materials, Cambridge, Massachusetts 02138, USA}
\affiliation{Department of Chemistry and Chemical Biology, Harvard University, Cambridge, Massachusetts 02138, USA} 
\date{\today}

\begin{abstract}
The ability to control electromagnetic fields on the subwavelength scale could open exciting new venues in many fields of science. Transformation optics provides one way to attain such control through the local variation of the permittivity and permeability of a material. Here, we demonstrate another way to shape electromagnetic fields, taking advantage of the enormous size of the configuration space in combinatorial problems and the resonant scattering properties of metallic nanoparticles. Our design does not require the engineering of a material's electromagnetic properties and has relevance to the design of more flexible platforms for probing light-matter interaction and many body physics.
\end{abstract}


\maketitle

The ability to manipulate electromagnetic fields could offer unprecedented opportunities in many fields of science, ranging from ultrasensitive biosensing\cite{a13}, data processing\cite{a14} to new platforms for probing many-body physics\cite{a15}. One way to obtain such control is through transformation optics\cite{b4,b5}. Enabled by the great freedom of design provided by artificially engineered structures known as metamaterials\cite{b1,b2,b3}, transformation optics has lead to a fruitful of exciting functionalities such as cloaking\cite{b2,b6}, materials with negative refractive index\cite{b1,b8}, perfect absorbers\cite{b3}, optical illusion\cite{c1} and metamaterial analog computing\cite{b7}. At the heart of transformation optics is the ability to locally vary the permittivity and permeability of a material.

\begin{figure}
\begin{center}
\includegraphics[width=7cm]{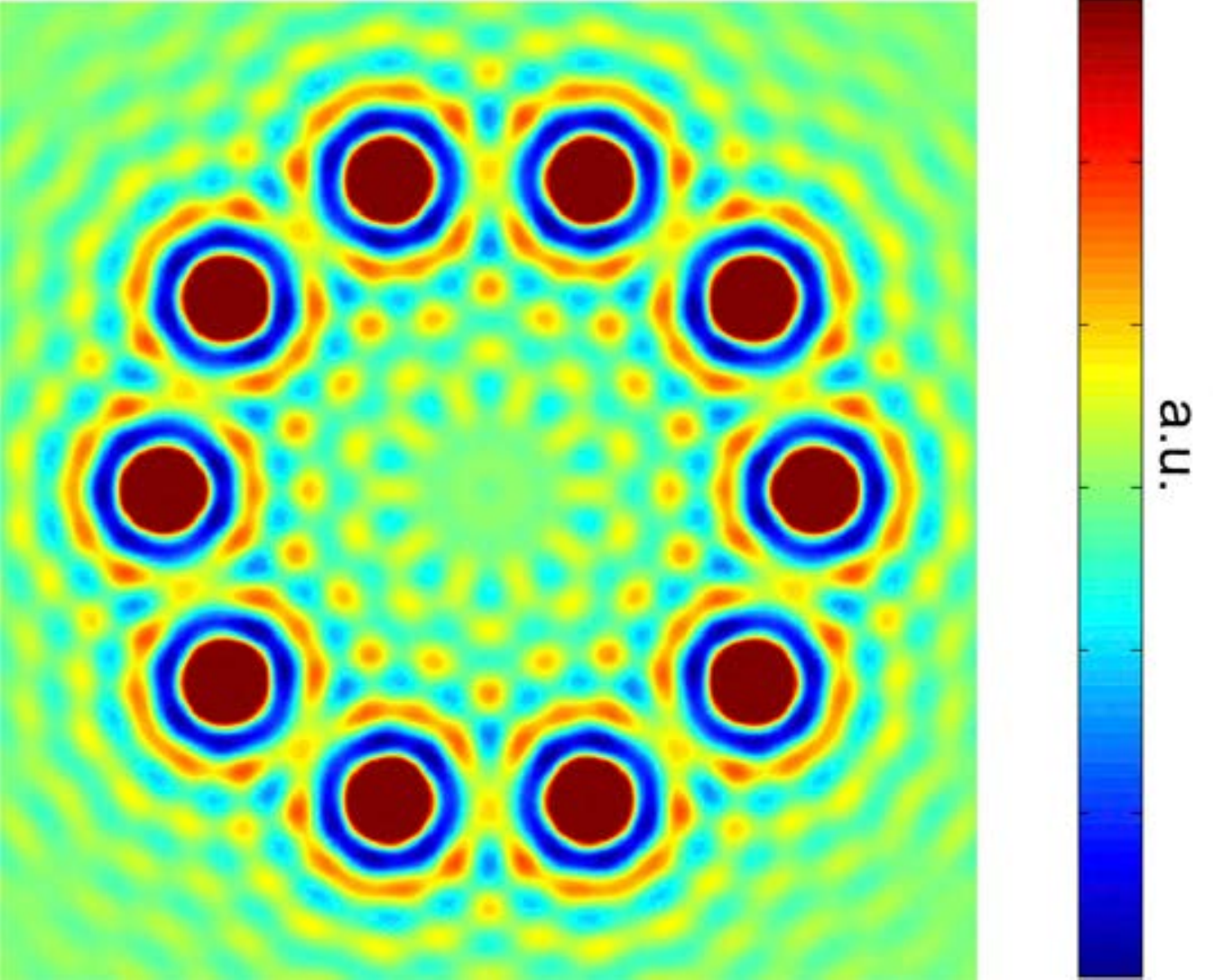}
\end{center}
\caption{\label{corral}\footnotesize{%(color online)
{\bf Plasmonic Corral.} The signal measured by a raster scanning s-SNOM tip when ten nanoparticles are uniformly arranged around the perimeter of a circle with radius $3\lambda$, where $\lambda$ is the wavelength of the incident light at the LSP resonance frequency.}}
\end{figure}

Here we demonstrate another possibility for controlling electromagnetic fields, without the need to engineer a material's electromagnetic properties. Our design takes advantage of the enormous size of the configuration space in combinatorial problems and the resonant scattering properties of metallic nanoparticles, which, when illuminated by light with the right frequency, give rise to resonant dipole modes known as the Localized Surface Plasmons(LSPs)\cite{a1}. Such LSP modes help bridge the gap between photonics and electronics\cite{a4,a5} and in many ways behave like the atomic/molecular point scatters used to build quantum corrals\cite{a2,a3}. In that case, the multiple scattering of the electron surface wave by the point scatters leads to "standing" wave patterns that could be probed by a scanning tunneling microscope(STM). The optical analogy of quantum corral was theoretically predicted\cite{a16} and experimentally probed using a scanning near field optical microscope\cite{a17}. 

For the optical corral, the arrangement of nanoparticles in space is given a priori and one only needs to solve a forward scattering problem to find out the resulting wave pattern. However, in order to shape electromagnetic fields, one instead needs to solve the inverse problem, for which the resulting wave pattern is prespecified and the configuration of nanoparticles needs to be determined. A simple calculation reveals the enormous number of possible configurations. For 80 nanoparticles with diameter 20nm arranged inside a two dimensional square with side length $6\lambda$($\lambda$ is the light wavelength at the LSP resonance frequency), the number of possible configurations is more than $10^{194}$, which is significantly larger even than the total number of atoms in the observable universe. This enormous configuration space present both difficulty and promise for a possible solution of the inverse problem.

We discuss two possible schemes for solving this inverse problem and in both cases, the control over the electromagnetic field is obtained through the manipulation of the locations of the nanoparticles, whose precise control has been enabled by the recent development in near field optical nanotweezers\cite{b7}.

\begin{figure*}
\begin{center}
\includegraphics[width=16cm]{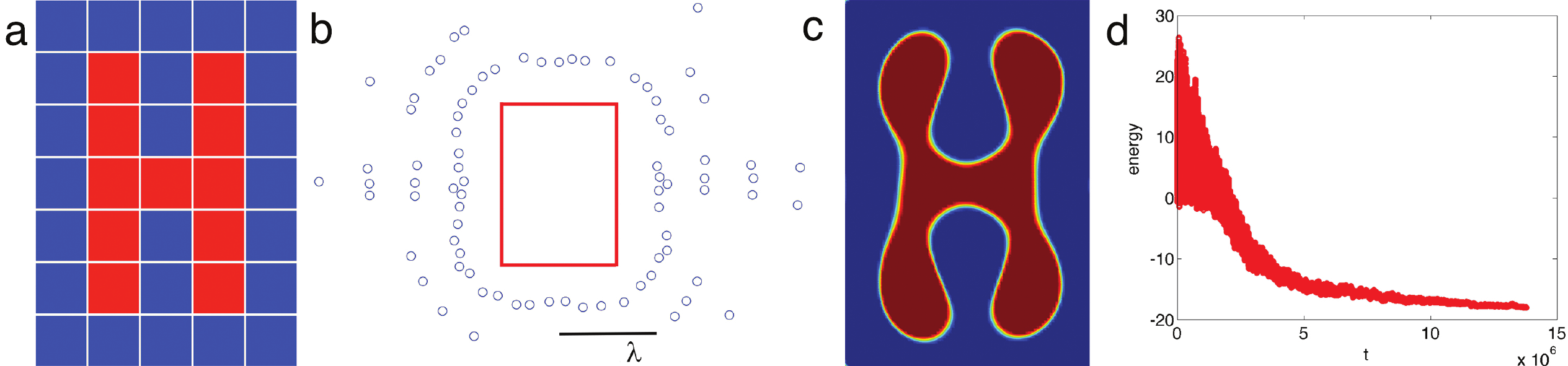}
\end{center}
\caption{\label{ien}\footnotesize{%(color online)
{\bf Plasmonic Hologram. (A)} The desired wave pattern discretized on a grid with grid spacing $\lambda/4$. {\bf (B)}The configuration of the nanoparticles that gives rise to the wave pattern "H". The black reference bar denotes one wavelength of the incident light. {\bf (C)} The wave pattern as measured by the s-SNOM tip when the nanoparticles are arranged as in (B). This whole region corresponds to the region inside the red rectangle in (B). {\bf (D)} The energy function at each time step. }}
\end{figure*}

The first setup closely mimics that for the quantum corral\cite{a2,a3} and can be thought of as a two dimensional plasmonic hologram. Gold nanoparticles are arranged on a two dimensional dielectric surface and a scattering type near field optical microscope(s-SNOM)\cite{a6,a9,a7,a8} tip is raster scanning across the surface while being illuminated by a focused laser beam at the LSP resonance frequency. In this setup, the tip acts both as an illumination source and a probe that reads out the optical signal. As a source, the tip sends out light in all directions, which then moves freely in space until being scattered by the gold nanoparticles. Some of the scattered light will be redirected back towards the tip and constitutes the signal being measured. As one moves the tip across the surface and records the signal at each position, it yields a two dimensional wave pattern. This physical picture is almost identical to the explanation for quantum corrals, except in that case, it is the surface electron wave\cite{a2} rather than light that is being scattered.

Our solution of the inverse problem relies on an efficient method to solve the forward scattering problem, for which we employ the same multiple scattering theory that explains the formation of both the quantum corral\cite{a2,a3} and the optical corral\cite{a16,a17}. In this approach, one first solves for the total electric field at each nanoparticle's position using the following self-consistent equations:
\begin{equation} 
\label{self}
\begin{aligned}
\vec{E}(\vec{r}_{i})=\vec{E}_0(\vec{r}_{i})+\sum_{j\neq i}G(\vec{r},\vec{r}_{i})\alpha(\omega)\vec{E}(\vec{r}_{j})
\end{aligned}.
\end{equation}
where $\vec{r}_i$ is the location of the ith nanoparticle(out of a total of N nanoparticles).  $\vec{E}_0(\vec{r})$ is the incoming field and $\alpha(\omega)$ is the polarizability tensor that characterizes the electromagnetic properties of the nanoparticles.

$G(\vec{r},\vec{r}')$ is the interaction tensor that describes how the electromagnetic field propagates in the absence of the nanoparticles. It includes contributions from both free space and the substrate. Thus, we can write it as a summation of two parts:
\begin{equation} 
\begin{aligned}
G(\vec{r},\vec{r}')=G_0(\vec{r},\vec{r}')+G_S(\vec{r},\vec{r}')
\end{aligned}.
\end{equation}

 $G_0(\vec{r},\vec{r}')$ is the interaction tensor in free space and is given by
\begin{equation} 
\begin{aligned}
G_0(\vec{r},\vec{r}')=(k^2+\bigtriangledown \bigtriangledown)\frac{e^{ik|\vec{r}-\vec{r}'|}}{|\vec{r}-\vec{r}'|}
\end{aligned},
\end{equation}
where k is the light momentum in free space. $G_S(\vec{r},\vec{r}')$ is the interaction tensor generated by the substrate alone and it is provided in Ref.\cite{a18}.

Once we find all the $\vec{E}(\vec{r}_i)$, the total electric field at any position $\vec{r}$ is then given by:
\begin{equation} 
\begin{aligned}
\vec{E}(\vec{r})=\vec{E}_0(\vec{r})+\sum_{i=0}^{N}G(\vec{r},\vec{r}_{i})\alpha(\omega)\vec{E}(\vec{r}_{i})
\label{wave}
\end{aligned}.
\end{equation}
This method is shown to agree with both experimental results\cite{a17} and more accurate numerical solutions\cite{d1,d2,d3} when the minimum separation between the nanoparticles is larger than twice their diameter, which we impose as a constraint in our algorithm.

\begin{figure*}
\begin{center}
\includegraphics[width=14cm]{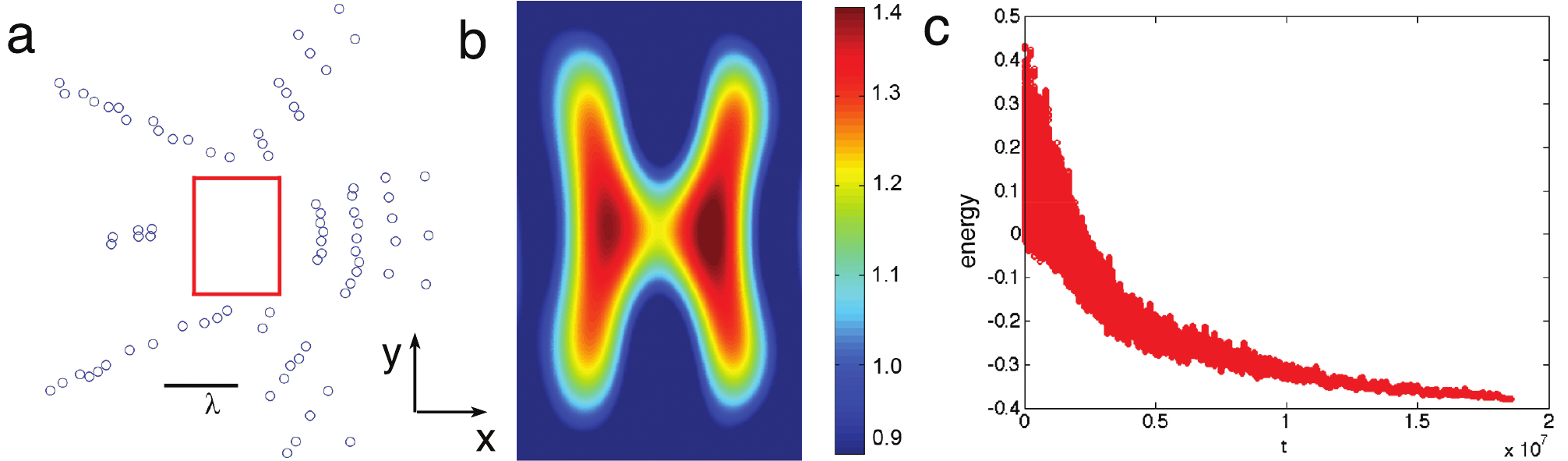}
\end{center}
\caption{\label{illum}\footnotesize{%(color online)
{\bf Engineering the illumination pattern. (A)}The configuration of the nanoparticles that gives rise to a scattering wave pattern resembling the letter "H". We have imposed a minimum spacing constraint on each nanoparticle's position in the algorithm, so no two nanoparticles are overlapping. The blue circles in this plot is significantly enlarged for better visualization.{\bf (B)} The scattering wave pattern($|E_z|/|E_0|$) that is generated given the configuration in (A) and a plane wave incident from the x direction. {\bf (C)} The energy function at each time step. }}
\end{figure*}

Once we solve (\ref{self}), we can plug the results into (\ref{wave}) to get the signal at the current tip position. This procedure is repeated for each tip position and a wave pattern is produced once the tip has scanned across the whole surface. In Fig.\ref{corral}, we show the resulting wave pattern when ten gold nanoparticles, each with diameter 20nm, are placed around the perimeter of a circle with radius $3\lambda$, where $\lambda$ is the wavelength of the incident light at the LSP resonance frequency.

For the inverse problem, similar problems were previously solved in the context of quantum mirages\cite{a19} and quantum corrals\cite{a3} using optimization methods like simulated annealing\cite{a20} and genetic algorithms\cite{a21}. Our approach is primarily based on simulated annealing, with the addition of an adaptive updating step. The simulated annealing algorithm is based on the annealing process in statistical physics, where a solid system is first melted at a high temperature and the temperature is then gradually lowered until it reaches the ground state.

Without loss of generality, we engineer a pattern that resembles the alphabetical letter "H". This letter "H" is first put onto a discretized grid as shown in  Fig.\ref{ien}(A), in which a red box represents a high signal while a blue box corresponds to a low signal. In order to implement simulated annealing, we first define an effective energy function that measures the "distance" between any wave pattern from the desired wave pattern. One convenient choice is 
\begin{equation} 
\label{energy}
\begin{aligned}
E=\max_{i\in Blue}m_i -\min_{j\in Red} m_j
\end{aligned},
\end{equation}
where $m_i$ is the signal measured at the center of the ith box. This energy function will be gradually decreased as one lowers the "temperature" of the system during the annealing process.

The system is initialized with a random configuration and a high enough temperature T that renders all possible configurations equally likely. Simulated annealing is then implemented in an updating step and an acceptance step. In the updating step, a random change to the current configuration is proposed, while in the acceptance step, the change in energy $\Delta E$ is calculated and the current configuration is replaced with the proposed configuration with the following probability:
\begin{equation} 
\begin{aligned}
P=\left\{\begin{matrix}
 1,&\Delta E\leqslant 0 \\ 
 exp(-\Delta E/T),& \Delta E>0
\end{matrix}\right.
\end{aligned}.
\end{equation}
This is the Metropolis-Hastings algorithm\cite{e1} and it leads to a Boltzmann distribution in equilibrium. After a target number of changes are accepted for a fixed temperature, the system is cooled down to a lower temperature\cite{a22} and the whole process repeats until a target temperature is reached or when no more changes are accepted.  

We slightly modify the updating step to incorporate information of low energy configurations that have appeared before the current time step. We divide the space into a finite number of regions and each region is given a frequency weight that determines how likely it is going to contribute a nanoparticle. Initially, all regions are equally weighted. At every time step, each nanoparticle is associated with a frequency weight equal to the value of a merit function of the current energy and this frequency weight is added to the region that nanoparticle belongs to. After a few time steps, this builds up a frequency profile in space where more frequent regions are more likely to contribute nanoparticles that yield low energy configurations. If one imagines each region as a gene, the merit function then measures the quality of a gene and more frequent regions correspond to genes with higher qualities. We use the following merit function for our simulations

\begin{equation} 
\begin{aligned}
f(E)=\left\{\begin{matrix}
 (1-\frac{E}{2T})^2,& 0<E<2T\\
min(exp(-E/T),exp(5)),& E\leqslant 0 \\ 
 0,& otherwise \\ 
\end{matrix}\right.
\end{aligned}.
\end{equation}

In the updating step, we randomly pick a nanoparticle from the current configuration. With probability p, we uniformly generate a random position for it and with the other 1-p probability, we generate a random position in its neighboring regions according to the current frequency profile.

The resulting configuration of the nanoparticles is shown in Fig.\ref{ien}(B) and the corresponding signal pattern is plotted in Fig.\ref{ien}(C). In these simulations, we use 80 gold nanospheres with diameter D=20nm and resonance frequency 2.2eV(corresponding to a wavelength $\lambda=563nm$)\cite{a23}. The s-SNOM tip is assumed to be polarized in the z direction. To guarantee the generality of the algorithm, we do not impose any symmetry constraint on the configuration of the nanoparticles. Interesting, a quasi-symmetric pattern emerges in Fig.\ref{ien}(b), which partially attests to the correctness of the results. When plotting Fig.\ref{ien}(c), our algorithm finds a threshold signal value $m_c$ such that when the signal measured at the current position is smaller than $m_c$, it is colored blue and red otherwise.

A plot of the energy at each time step is shown in Fig.\ref{ien}(D). As one can see, the energy on average decreases over time with diminishing fluctuations. The average energy for random configurations of nanoparticles is 11.9(arbitrary unit) with standard deviation 4.78. The final energy for the configuration in Fig.\ref{ien}(c) is -18.03, which is more than six standard deviations away from the average. Even though it is no guarantee that we have found the global minimum\cite{a20,a22}, such a large gap is good enough for most practical purposes.

In the above case, the s-SNOM tip acts both as a signal probe and an illumination source. Since the source itself is moving, the signal being read out is not the scattering wave pattern, but a result of the interference between the incident wave and the backscattered wave at each tip position\cite{a2,a3,a6,a7,a16,a17}. 

A possibly more interesting problem is to engineer the illumination pattern itself, that is, to engineer the scattering wave pattern given a fixed incident wave.  Subwavelength control of this illumination pattern can be relevant for a wide range of scenarios, including but not limited to the design of more flexible platforms for probing many body interaction\cite{a15,a24,a25,e2,e3}.

Consider an incident plane wave of the form $\vec{E}_0(\vec{r})=E_0e^{ikx}\hat{z}$. The same multiple scattering approach applies with a modified incident condition. The results are shown in Fig.\ref{illum}. In this case, the average energy(normalized by the magnitude of the incident wave) for completely random configurations is 0.2 with standard deviation 0.06. The final energy for the pattern in Fig.\ref{illum} is -0.38, which is 9.7 standard deviations away from the average.

We have shown that one can gain a considerable amount of control over the shape of electromagnetic fields using metallic nanoparticles. Our scheme does not require spatial or temporal engineering of the electromagnetic properties of a material and can be employed to engineer  electromagnetic fields in free space. Such design has relevance to the design of more flexible platforms for probing light matter interaction or many body effects.

We thank Harvard Institute for Applied Computational Science for generous financial support and the results are produced using Harvard FAS high performance computing system Odyssey. EJH was supported by U.S. Department of Energy under Grant DE-FG02-08ER46513 and the STC Center for Integrated Quantum Materials under NSF Grant No. DMR-1231319.

\end{document}